# Musico-acoustic Depictions of Laminar and Turbulent Flows in Ligeti's Piano Étude No. 9 and a Novel Method of Analysis

**Noah Chuipka**
**April 26, 2022**

## Abstract

The relationship between musical material and physical phenomena has become a topic in the musicological literature over the last several decades, particularly concerning elements of the musical system itself, and constructions found in the work of contemporary classical composers such as György Ligeti and Iannis Xenakis. Most scholars, who adopt this approach, explore the physical phenomena of fractals in the analysis of musical works, but fluid mechanical frameworks, such as laminar and turbulent flows, offer a new avenue to be explored. In this paper I will propose a novel method of musical analysis for examining musical structures in terms of fluid-like behaviour such that Ligeti's étude no. 9 serves as a model, whereby the metaphors of laminar and turbulent flows take precedence. The methodological design includes the utility of converting terms (by proposing correlations between physical concepts and the acoustic properties of music), theoretical frameworks for musicological application, and scatter plots, which provide central analytic support to demonstrating the fluid-like tendencies in musical materials, for they capture a formal development over time.

> *Although I am an artist, my working method is that of a scientist active in basic research rather than in applied science. Or of a mathematician working on a new mathematical structure, or of a physicist looking for the tiniest particle of the atomic nucleus. I do not worry about the impact my music will make or what it will turn out to be like. What interests me is to find out the way things are. I am driven by curiosity to discover reality. Of course, there is no reality in art the way there is in science, but the working method is similar. Exactly as in basic research where the solution of a problem throws up innumerable new ones, the completion of a composition raises a host of new questions to be answered in the next piece.* - Ligeti, G (Varga, 2013, p. 32).

## 1 Introduction

Claims of mathematical and scientific depictions in Ligeti's work have emerged in the musicological literature, but mostly with regards to geometrical concepts vested in nature, mathematics, and art. For instance, the proposed fractals in the 4$^{th}$ movement of Ligeti's Piano Concerto (Steinitz, 1996a); the repeating series of expanded and contracted pitch structures in étude no. 14 that depict the repeating ascending columns in Constantin Brancusi's (1876-1957) 29-metre-high sculpture (Steinitz, 1996a); appearances of Lorentz's butterflies, Koch's curve, Gaston Julia's fractal, and Cantor's function in the études (Blanaru, 2020); and chaotic determinisms in étude no. 1 (Steinitz, 1996b), where initial musical patterns (i.e., initial conditions) result in similar yet dispersive patterns. Although most of the relevant literature discusses geometry and chaos, Ligeti's own words suggest that he was also interested in topics such as fluid mechanics—to which analyses of particular works can transpire evident musical resemblances to those familiar with the behaviour of fluids.

Describing music in terms of physical motion is not uncommon, but is rarely fully articulated as a direct conversion from physical phenomenon to music; that is, in a more tangible way that accommodates both musical and physical descriptions. In his 2002 study, Larson argues that we often metaphorically interpret musical motions as physical motions; for his study, he focuses on three "musical forces" that may be realized:

> "gravity" (the tendency of a note heard as "up in the air" to descend), "magnetism" (the tendency of an unstable note to move to the nearest stable pitch, a tendency that grows stronger the closer we get to a goal), and "inertia" (the tendency of a pattern of musical motion to continue in the same fashion, where what is meant by "same" depends upon what that musical pattern is "heard as"). That is, even though music does not literally move, experienced listeners of tonal music hear that music metaphorically as purposeful physical action—subject to musical forces that are analogous to the physical forces that shape physical motion. (p. 352)

Metaphors provide a useful way of drawing connections between physical forces and musical experience. This metaphorization may be expanded to include the physical phenomenon of fluid motion in the context of a musical experience. Upon listening to and analyzing the works of Ligeti—but more particularly, his études—we may conclude that many of the phrase structures closely resemble fluid flows, a concept in fluid mechanics. Fluid mechanics involves the study of the behaviour of fluids in both resting and moving states, whereas fluid dynamics—that which we will be more concerned with here—comprises the study of fluids in only moving states. Physicists agree that fluids include any physical material capable of flowing; namely, gases and liquids. A knowledge of both fluids and advanced music theory allows one to associate *pitches in motion* with *fluids in motion*. Although metaphors provide the basis for this association, the acoustic properties of music cause particularly more direct correlations between the fluid dynamical and the musical.[1]

## 2 Methodology

I will analyze the musical behaviour of Ligeti's étude no. 9 to examine the laminar and turbulent resemblances by which multiple extramusical methods can be adequately applied. First, principal fluid mechanical concepts undergo conversion to principal musical properties—this serves as a basis for the 'musico-physical' framework, and reveals the capacities for a musicological method of analyzing music through the perspective of fluid behaviour. The proposed conversions elucidate a number of analytic tools that mathematically (e.g., time series) and intuitively correspond to musical properties, however the features of dynamics required an adaptation where a collection of incremental natural numbers represent musical dynamics. To effectively describe laminar and turbulent flows with the fluid-to-music conversions, we require a propositional set of formal conditions. Lastly, a computational framework native to physics—

[1] Notably pressure, discussed below in section 2.

scatterplots—is constructed to demonstrate how the time and pitch placement of a musical structure presents trends correlative to the time and particle location of a physical event. Table 1 presents a summary of the four extramusical methods employed.

| Extramusical method | Praxis |
|---|---|
| Conceptual conversions | Correlates fluid to musical features. |
| Numerical representation of dynamics | Correlates numerical growth to dynamical levels. |
| Conditions | Proposes a set of formal conditions to describe a laminar and turbulent flow. |
| Scatterplots | Demonstrates temporal trends in music that correspond to spatiotemporal trends in physics. |

**Table 1.** Summary of extramusical methods.

## 3 A Brief Review of Fluid Mechanical Concepts

To begin, table 1 and its corresponding visual representations provide selective basic definitions in fluid mechanics—which includes fluid dynamics as a subdiscipline—that are crucial for our discussion. These include primarily concepts related to force and motion. From hereon, we will merely use the term 'mechanical' in the context of fluid mechanics.

| **Term** | **Definition** |
|---|---|
| Fluid particle | A small unit of matter that has fluid properties. |
| Flow | The motion of a fluid particle. |
| Layer | An arrangement of fluid particles. |
| Inertia | The tendency to remain in uniform motion, or in a resting state. |
| Resistance | The measure of an opposing force. |
| Inertial force[2] | The force that represents equal and opposite reaction of the fluid. |

[2] Also known as *fictitious force*.

| | |
|---|---|
| Viscous force | The force between fluid layers that oppose relative motion. [This is used to measure the resistance to a change in flow.] |
| Flow field | A space of physical quantities in which fluid particles exist. |
| Adjacency | When two fluid pathlines are next to one another (see fig. 5). |
| Velocity | Speed over time in a given direction. |
| Pressure | The application of force against an object. |
| Density | Mass per unit of volume. |
| Viscosity | The fluid's tendency to oppose the relative motion between its layers. |
| Characteristic linear dimension | The physical dimension (often length) that contains the flow.[3] |
| Timeline | A set of adjacent fluid particles that form a line at a given instant. |
| Pathline | The trajectory of a fluid particle in motion. |
| Streakline | A line that marks the area of a flow of which fluid particles have passed. |
| Streamline | Lines in the flow field that are tangent to the direction of the flow at every point at a given instant. |
| Laminar flow | Parallel adjacent fluid layers, with no disruption between the layers. |
| Turbulent flow | A fluid motion that experiences chaotic changes in magnitude and direction. |
| Boundary layer | A layer of fluid that is closest to a surface. |
| Smoothness | The linear motion of adjacent layers. |

**Table 2.** Fluid Mechanical Concepts.

### 3.1. Laminar and turbulent flows

Laminar and turbulent flows (fig. 1) commonly occur in both natural and engineered constructions. The parallel layers in a laminar flow are often described as smooth with respect to their linearity, and the smoothness is lost often through the anomalies of mixing between layers

[3] In conduit pipes—which is a common domain for fluid dynamical modelling—the diameter of the pipe is considered.

(i.e., directional change), a change in velocity to a higher magnitude, and a change in pressure to a higher magnitude. Importantly, laminar flows always precede turbulence, and the transition to turbulence has been identified as the *laminar-turbulent transition*. A number of theories present the varying ways ('phases') in which a laminar flow transitions to a turbulent flow, but the general initial conditions for a transition consist of environmental disturbances in the flow profile such that changes occur in frequency, pressure, direction, and velocity. The changing properties in a transition are small-scale instabilities, whereby physicists study laminar-turbulent transitions primarily to both predict the potential occurrence of large-scale instability (i.e., turbulence) and to understand the varying properties that characterize the transitions themselves.

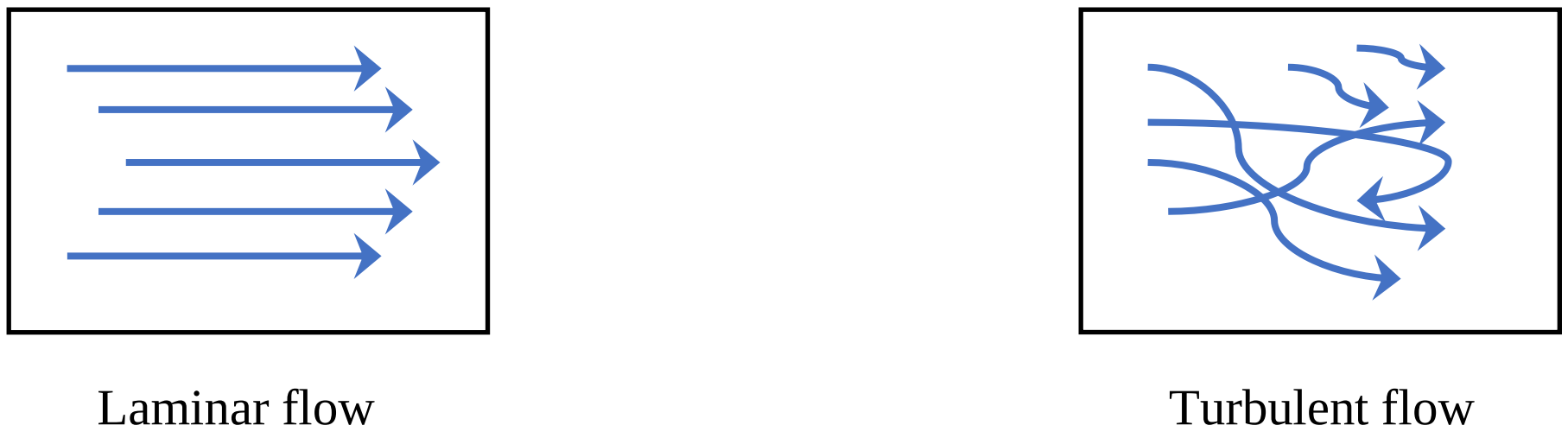

**Figure 1.** Laminar and turbulent flows.

Notable examples of laminar-turbulent flows include waterfalls and candlelight smoke (see fig. 2 and 3). During laminar-turbulent transition phases, the flow maintains laminar properties, but at the expense of turbulent *spots*[4]: noisy (i.e., an anomaly from the pattern) segments in the orderly flow. In simple terms, one can regard the laminar-turbulent transition as the link between balance and imbalance, whereas the link contains aspects of *both* balance and imbalance.

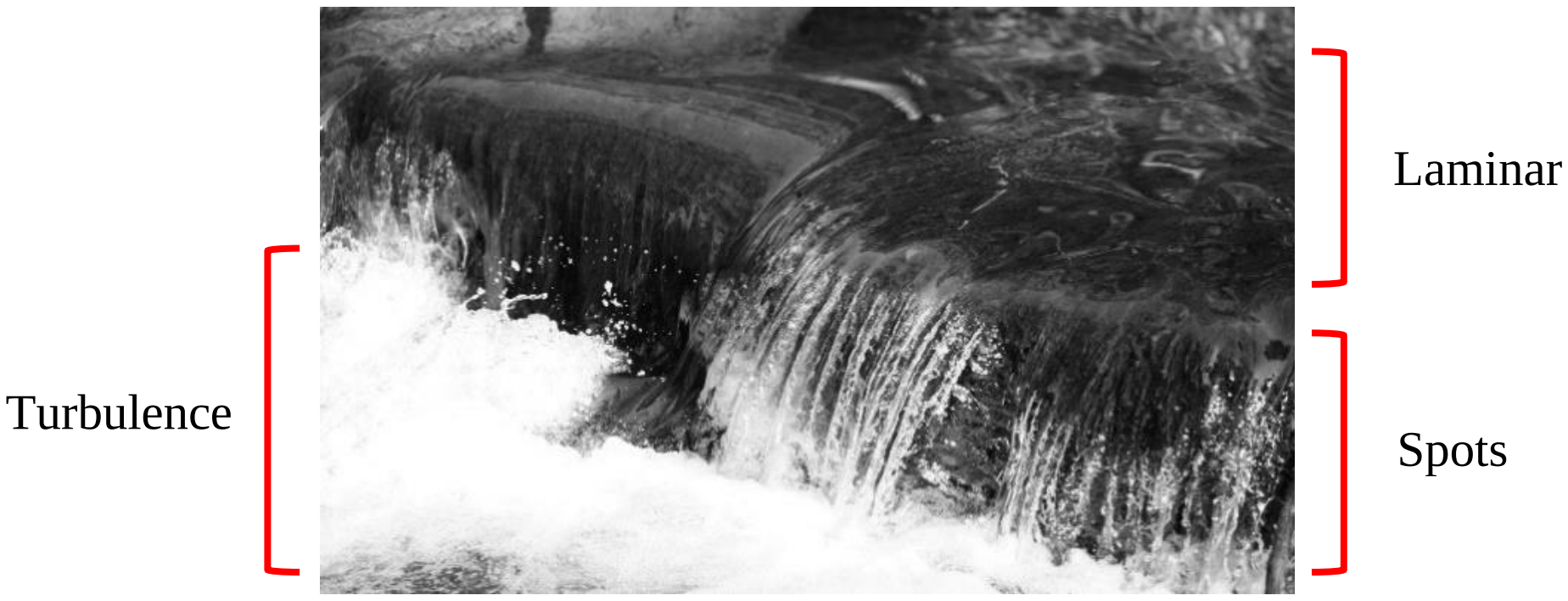

**Figure 2.** A waterfall demonstrating laminar-turbulent flow.

Note. From Vines, J. (2008). Sandvedparken. [Photograph].

[4] Sometimes referred to as "turbulent bursts."

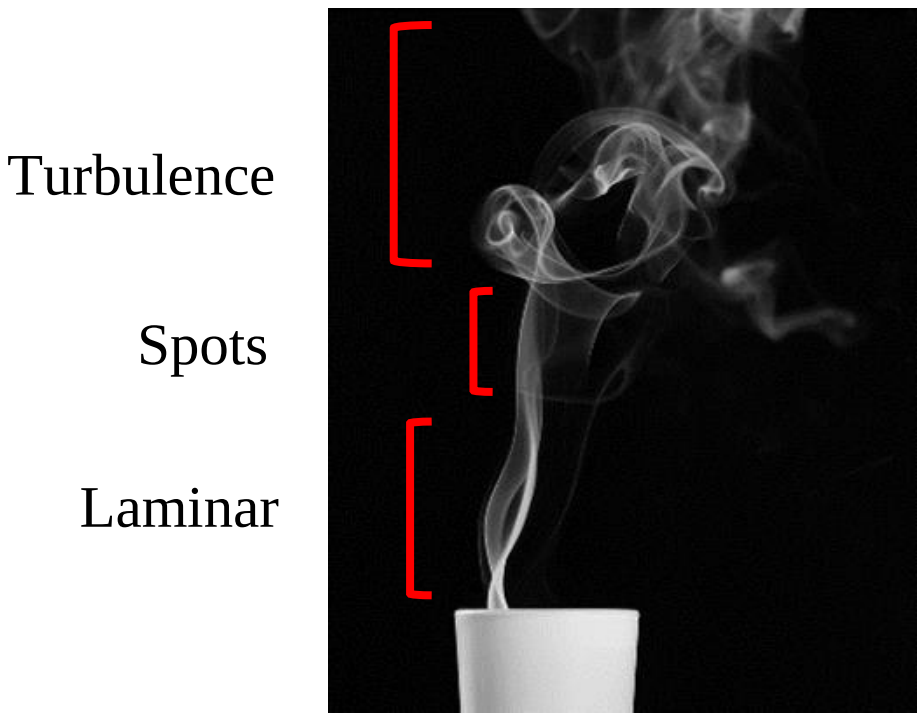

**Figure 3.** Candlelight smoke demonstrating laminar-turbulent flow

Note. From Winkler, F. (2021). [Image].

### 3.2. Fluid Dynamical to Musical Conversion

Table 3 below details the proposed correspondences between fluid mechanical and musical properties. Since we are concerned with the *movement* of particles and pitches, and not their specific properties, only mechanical concepts rooted in motion are applied to our musical framework, whereas those independent of motion (for e.g., mass) will be excluded. Accordingly, the list of conversions is not exhaustive to all pertinent fluid dynamical concepts—though the necessary concepts for defining laminar and turbulent flows receive centrality. Conversions which are not necessary for our framework are identified as N/A for *not applicable* in the "Musical" column, and the rows highlighted in blue designate the concepts necessary for the musical fluid analysis below.

| **Mechanical** | **Musical** |
|---|---|
| Fluid particle | A pitch.[5] |
| Flow | Pitches in motion, i.e., rhythm. |
| Inertia | The expectation of a pattern's or pitch's continuation. |
| Inertial force | The listener's mental assignment of opposing motion, with the condition that an inertia is cognized. |
| Viscous force | The pattern's or pitch's resistance to continue in the expected direction. |

[5] Since a particle is the principal unit of independent form in fluid mechanics, then a pitch, which too is the principal unit of independent form in music, aligns well with its mechanical counterpart.

| | |
|---|---|
| Flow field | A specified set of pitches. |
| Adjacency | Harmonic interval(s), i.e., when two trajectories are next to each other (see fig. 4). |
| Trajectory | A pitch that ascends (upwards motion) or descends (downwards motion). |
| Velocity | Fixed rhythm of which the pitches proportionately ascend or descend.[6] |
| Pressure | Dynamics. |
| Density | The quantity of harmonic layers. (The more layers, the higher the density.) |
| Viscosity | N/A. |
| Characteristic linear dimension | N/A. |
| Parallel flow | Two or more musical lines in parallel contrapuntal motion. |
| Timeline | A set of harmonic intervals at a given metrical time (i.e., beat). As in fluid dynamics, $t_n$ will represent a given beat.[7] |
| Pathline | The trajectory of a pitch. |
| Streakline | A line that marks the trajectory of a timeline. |
| Streamline | N/A |
| Boundary layer | N/A |
| Smoothness | Chromatic (smallest interval in equal temperament)[8] |

**Table 3.** Fluid Dynamical to Musical Conversion.

Since the intensity of musical dynamics lacks a quantitative assignment, an adaptation for quantities as pressure ($Pa$) magnitudes (table 4) will assist our analytic descriptions. To adapt, incrementing natural numbers assigned to increasing pressure magnitudes represent respective dynamical markings. Instead of ranging from the more standard *ppp* to *fff*, an analysis of Ligeti's work requires his notational convention of a *pppppppp* to *ffffffff*[9] range.

[6] While it would be plausible to include musical “velocity” for repeated pitches over a fixed rhythm, it would not produce the impression of movement (or *flow*) in a musical sense.
[7] See appendix of symbols.
[8] According to Straus (2003), for instance, smoothness in voice leading may be described as how voices travel the “shortest possible distance”. Although Strauss (2003) was referring to the syntax of pitch-class space voice leading, such a notion of smoothness can be equally applied, for the “shortest possible distance” that corresponds to our model of fluid dynamical is the distance between registral pitches (i.e., chromatic intervals).

| Dynamic | Number representative of pressure magnitude |
|---|---|
| *pppppppp* | 1 |
| *ppppppp* | 2 |
| *pppppp* | 3 |
| *ppppp* | 4 |
| *pppp* | 5 |
| *ppp* | 6 |
| *pp* | 7 |
| *p* | 8 |
| *mp* | 9 |
| *mf* | 10 |
| *f* | 11 |
| *ff* | 12 |
| *fff* | 13 |
| *ffff* | 14 |
| *fffff* | 15 |
| *ffffff* | 16 |
| *fffffff* | 17 |
| *ffffffff* | 18 |

**Table 4.** A conversion for musical dynamics to enumerated magnitudes of pressure.

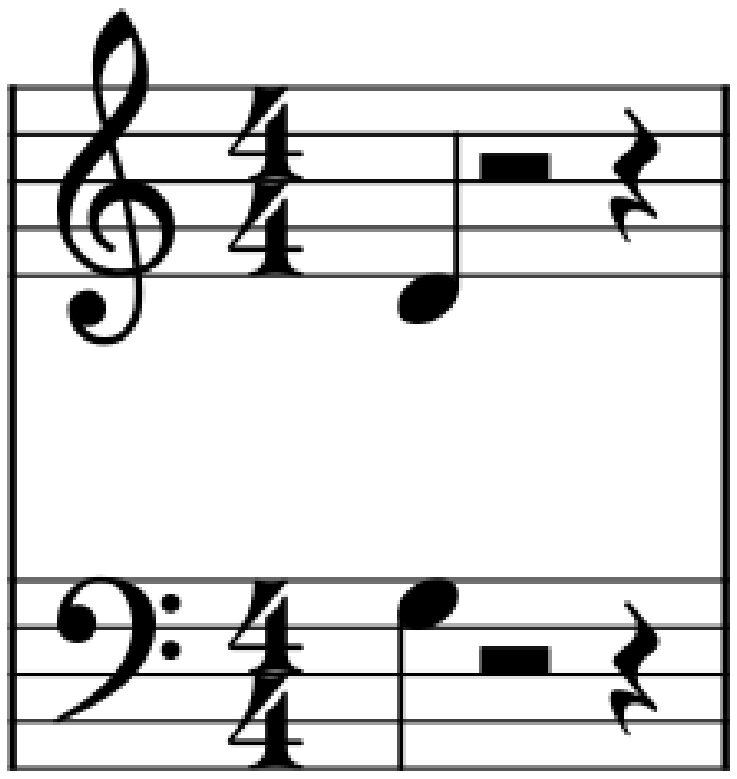

**Figure 4**. Vertical pitches.

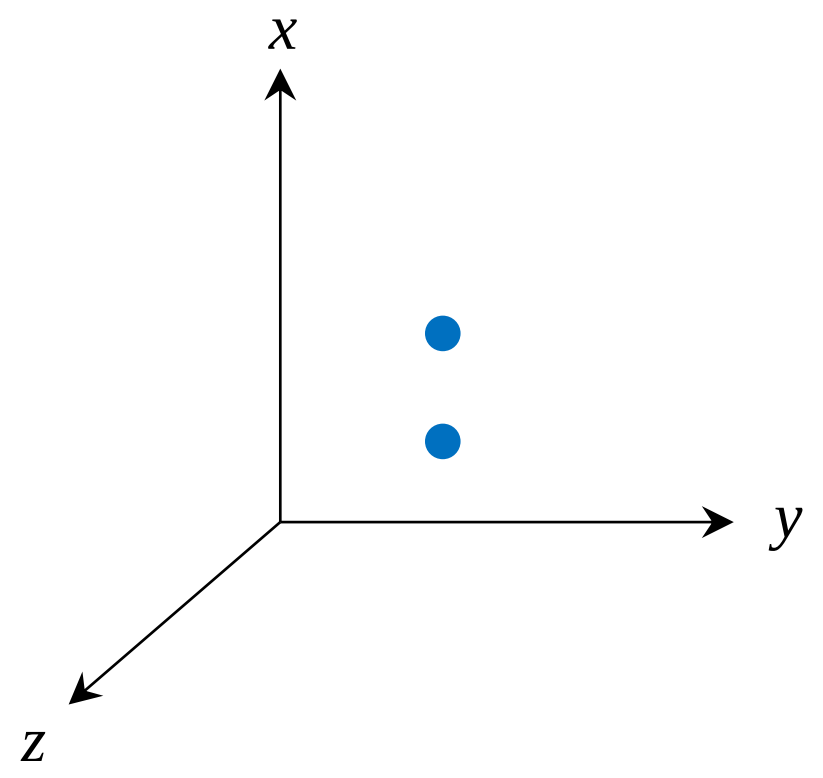

**Figure 5**. Adjacent fluid particles in Euclidean space.

[9] For example, see Ligeti's étude no. 9 for the lowest and no. 13 for the highest.

### 3.3. A Discussion of Selective Conversions

While most of the relationships between the mechanical and musical concepts may be intuitive, three are not as straight-forward and require more explanation before being applied to a musical work. These concepts consist of inertia, inertial force, and viscous force.

Inertial force defines the type of force resultant from an application of external force upon an object that demonstrates resistance against a change in velocity.[10] For example, if you push a box slowly, you can tactilely feel your own force exerted on the box through its resistance. Considering Newton's third law of action-reaction, inertial force characterizes the behaviour of an equal and opposite reaction—these two forces acting upon each other create a balance in physical behaviour. To define inertial force in a musical context, we must be clear on a definition of musical inertia. Prior to proposing a musical model, it is important to question what the inertia of a pitch might be. The simplest answer entails that the inertia of a pitch fails to exist, for if a human plays a pitch on an instrument, no physical force causes the occurrence of a successional pitch—there is thus no action-reaction with respect to the direction of a pitch. However, Larson's (2002) theory of musical forces, introduced earlier, granted an adequate analogy for this 'problem'—the conception of a momentum formed by our memory, where a musical pattern is expected to continue by the same means it began. Although Larson's reference of 'pattern' often signifies structures of tonal voice leading, he describes the concept of inertia in a general fashion that allows interpretations based on any musical motion capable of prediction—and, his focus on structures (i.e., 'patterns') rather than individual pitches does not necessarily limit the possible interpretations as well, for a listener may expect a singular resonating pitch to either move up, down, or repeat (i.e., that the pitch has 'momentum'). Accordingly, the consideration of inertia as the listener's expectation of a pattern's or pitch's continuation, provides the theoretic basis to musically conceive inertial force: the listener's mental assignment of opposing motion, with the condition that an inertia is cognized. Following closely to this, viscous force presents the concept that a pitch or pattern possesses some intramusical resistance to continue in the expected direction.

### 3.4. How to define a laminar and turbulent flow in musical terms?

To define a turbulent flow, we require the notion of a laminar flow in a musical sense. For a musical laminar flow, the musical material must meet the following proposed conditions: *a*) the music must include at least two lines adjacent to each other in parallel motion, *b*) the adjacent lines must share the same dynamic, and *c*) the adjacent lines must share the same rhythm. These three conditions will be identified as the L-parameters (laminar parameters) for the sake of concision. The latter two conditions (*b* and *c*) are necessary for defining turbulence, for a fluctuation in the constancy of pressure entails *b*, and a fluctuation in the constancy of velocity entails *c*. Hence, to define a musical turbulent flow, two conditions must be met: *a*) a change in magnitude (namely velocity and pressure), and *b)* a change in direction (i.e., contrapuntal motion). We will label these the T-parameters (turbulent parameters), and note that they are simply a variation in conditions established by the L-parameters. Importantly, we must highlight that, although we could require a musical laminar flow to be harmonically equivalent—for instance, if a pathline begins with major thirds, then it can only continue in major thirds—the homophony would not be as interesting as polyphony.

[10] In physics, pressure and velocity are related, but these are distinct in music. For instance, higher velocity entails higher pressure in physics, but not in music. Despite this divergence, both mechanical concepts will be related in the music by modifying musical parameters.

### 3.5. Transitional timelines

Although the inclusion of the laminar-turbulent transition phase would permit the full application of fluid mechanics to the musical framework, many of the physical properties are inapplicable in relation to music. Foremost, the notion of a boundary layer, which holds the laminar-turbulent transition together, is not transferrable to acoustic phenomena (and thereby, musical motion). We will thus select a few descriptions of mechanical behaviours—that can function notwithstanding the notion of a boundary layer—of the theories, and omit the others. The musical adaptation of these phases can be quantified in two categories: a) gradual increments in velocity and pressure, and b) a change in direction of a layer simultaneous to a constant direction[11] (fig. 6 and 7). Hence, there are two categories of which musical laminar-turbulent phases can occur: in graduation and in layer (see table 5). In directional change, the disruption of a layer's motion can also be accounted for by the cessation (i.e., a decay of the constancy) of an identical motion rather than merely changing the motion from up to down, or down to up (fig. 7).

| **Graduation** | **Layer** |
|---|---|
| Velocity | 'Spots' of directional change. |
| Pressure (dynamics) | |

**Table 5.** Musical aspects of laminar-turbulent regions.

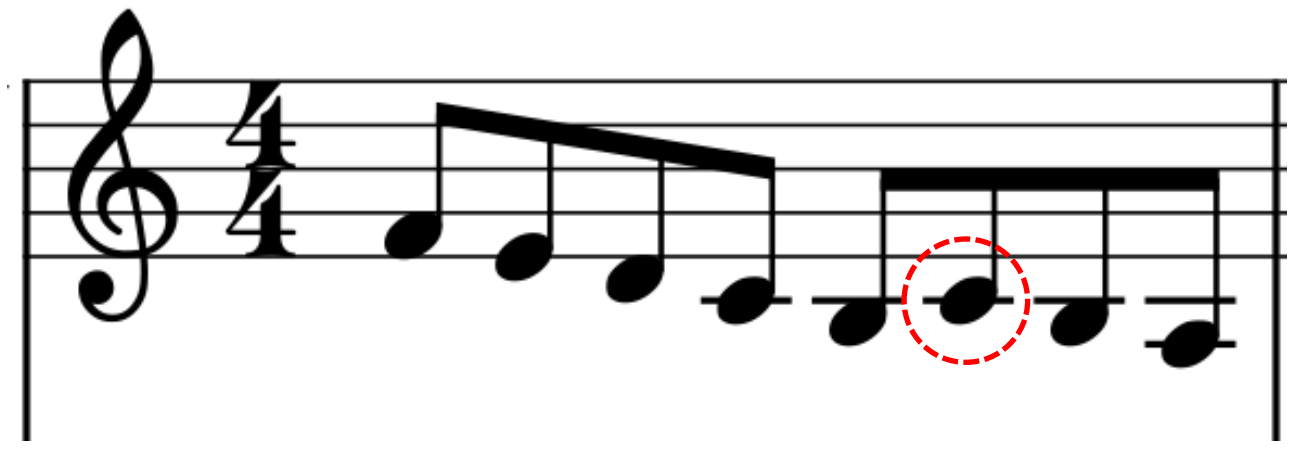

**Figure 6.** An example of a layer's turbulent spot where the downwards direction changes to upwards. The red circle outlines the change in direction.

[11] Hence, if the musical motion descends only, then a disruption to its potential continuation will result in a change.

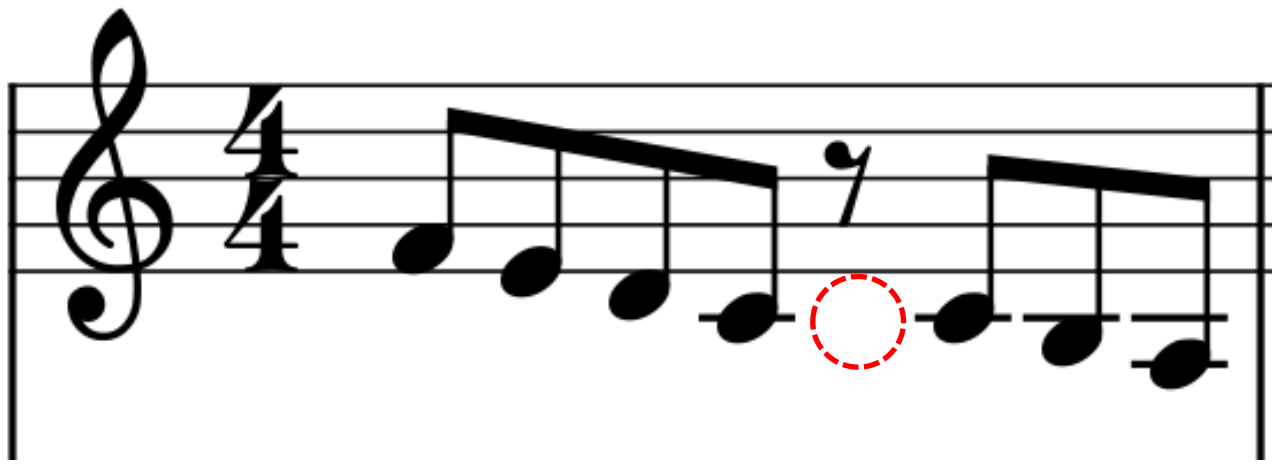

**Figure 7.** An example of a layer's turbulent spot where the downwards direction halts. Red circle outlines the change in direction.

## 4 Étude No. 9 (Application)

Besides the evident fluid-like material of smooth and mixing layers in his ninth étude, Ligeti assigned denotations that hint at the depiction of a fluid flow. Ligeti noted, regarding tempo, "So fast the individual notes—even with pedal—almost melt into continuous lines" (Ligeti, 1998), and for rhythm, "The piece has no rhythmic metre—it consists of a continuous flow ..." (Ligeti, 1998). *Continuity*, *flow*, and *melting* are all key concepts in fluid dynamics. Additionally, the use of legato—of which Ligeti indicates *sempre molto legato*—contributes to the notion of laminar layers being "smooth", for the meaning of legato is also described as being characterized by smoothness. Even if this étude was not intended to depict segments of laminar and turbulent flows, we can plausibly assume that—given his known infatuation with mathematics and physics—Ligeti's thought process involved depictions of fluid.

### 4.1. An Opening of Laminar Layers

A first glance at this étude leads the eye to Ligeti's extensive use of crossing beams. The first beam groups a descending chromatic scale in a constant rhythm and dynamic from a B to a lower octave A♭ at the beginning of the étude (see fig. 8). This is the initial (red) layer, and its limit is defined by the end of the chromaticism (i.e., smoothness).[12] Simultaneous to the initial layer, a new layer (blue) begins a minor sixth above such that the layer onset is identical to the initial layer onset. The new layer causes a laminar flow in satisfying all three of the L-parameters for the two layers in question; that is, the simultaneous movement of both layers are parallel, and identical in velocity and pressure. A third (green), then fourth (purple), layers enter the flow in an identical fashion. This harmonic syntax continues throughout the rest of the étude, but with superimposed phrases. Significantly, this opening material demonstrates the complex construction of descending chromatic lines that reset while maintaining continuity in the same register. Ligeti undoubtably sought to visually present this structural aspect by his apparent use of extensive crossing beams—likely to promote both a structural understanding as well as a performative maintenance of the continual melodic lines.

[12] The beam quantifies the length.

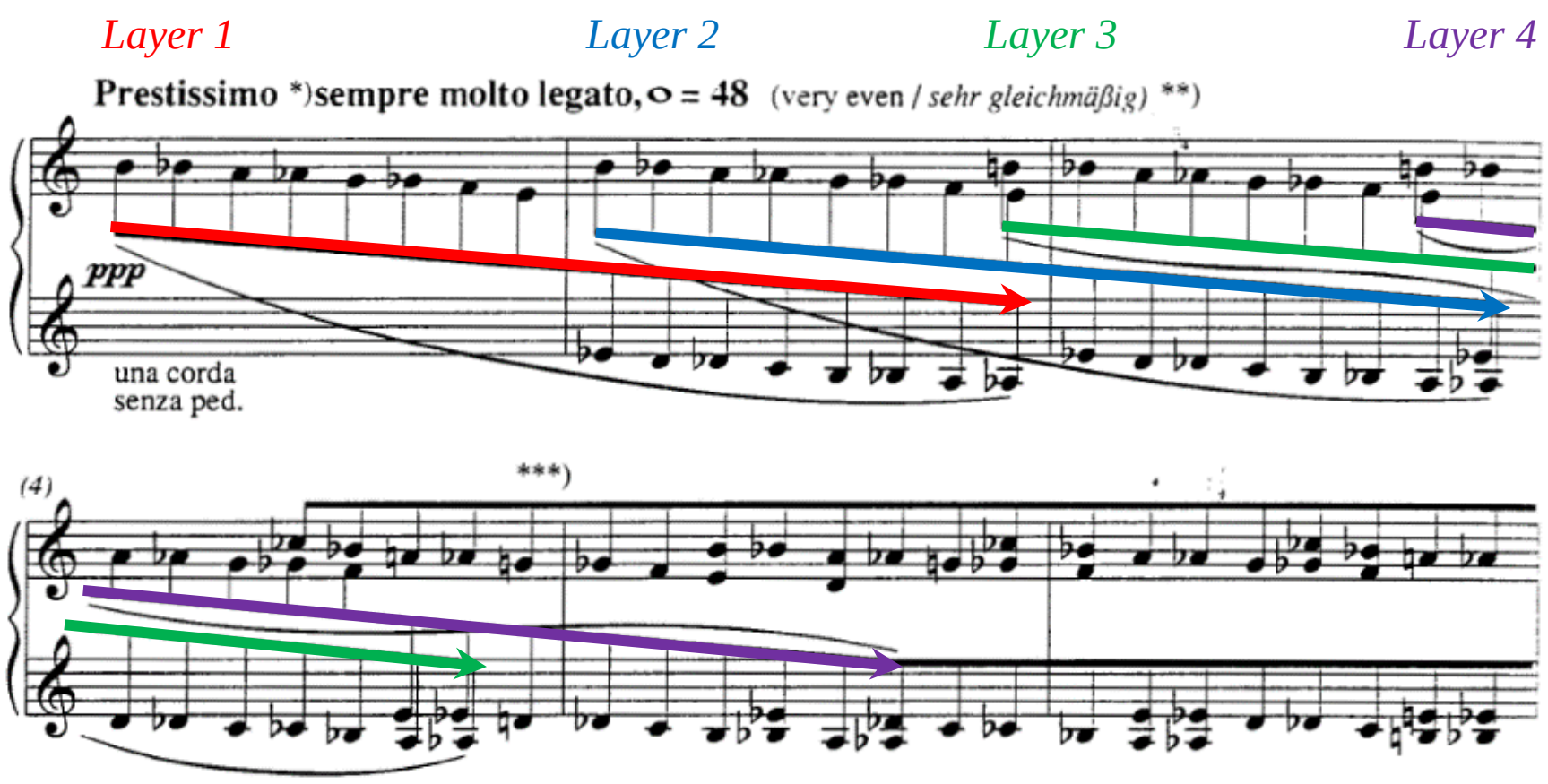

**Figure 8.** Étude No. 9, mm. 1-6. Outline of the first four layers.

Now that we have defined the elementary fluid properties of the first several layers in the étude, we can begin to represent musical motion. As stated in table 3, a timeline $t_n$ enumerates a single metrical unit; however, given the continuity of fluid, measure lines must be removed[13], and thus, the beginning of a composition yields $t_n$ and linearly continues until the last harmonic interval of the composition. Figure 9 outlines the musical particles and timelines for mm. 1-5, however, one will notice that the illustration in figure 9 does not adequately capture the constant downwards linearity of the flow layers—especially due to the enharmonic equivalents. The illustration serves as a basis to demonstrate how I will hereon identify the fluid layers: by the manner in which they begin at a highest point, and chromatically descend to a lowest point (whereby the descension no longer continues from that given lowest point).

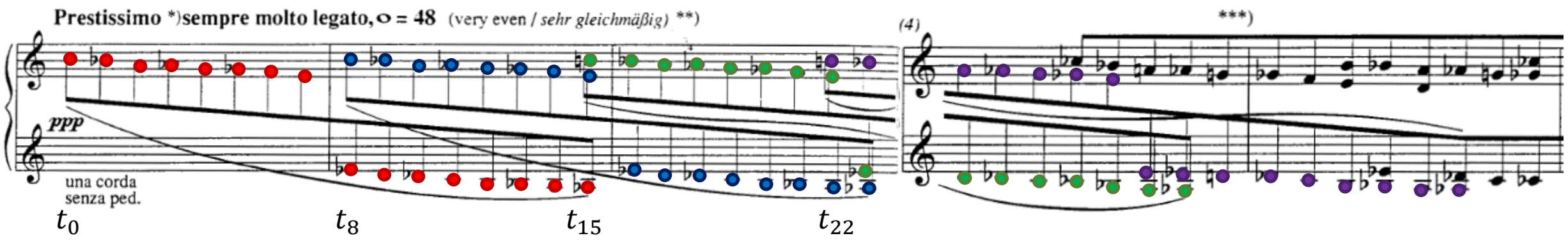

**Figure 9.** Étude No. 9, mm. 1-5. Particles and timelines of the first four layers.

Figure 10 reproduces Figure 9 without staff lines to show more clearly the musical motion as fluid flow. A collection of pitches from the first three measures are illustrated (in proportion to their placement on the staff) as a pathline of fluid particles over Euclidean space[14] (fig. 10). Although not as important for our theoretical framework, this provides a useful means for musicologists or performers to discern the patterns more easily—which, as mentioned above, are supported by Ligeti's crossed beams.

[13] Intriguingly, Ligeti often disregarded measure lines in search of continuity (for example, see Ligeti's description in études no. 3, 7, 12, and so on.).

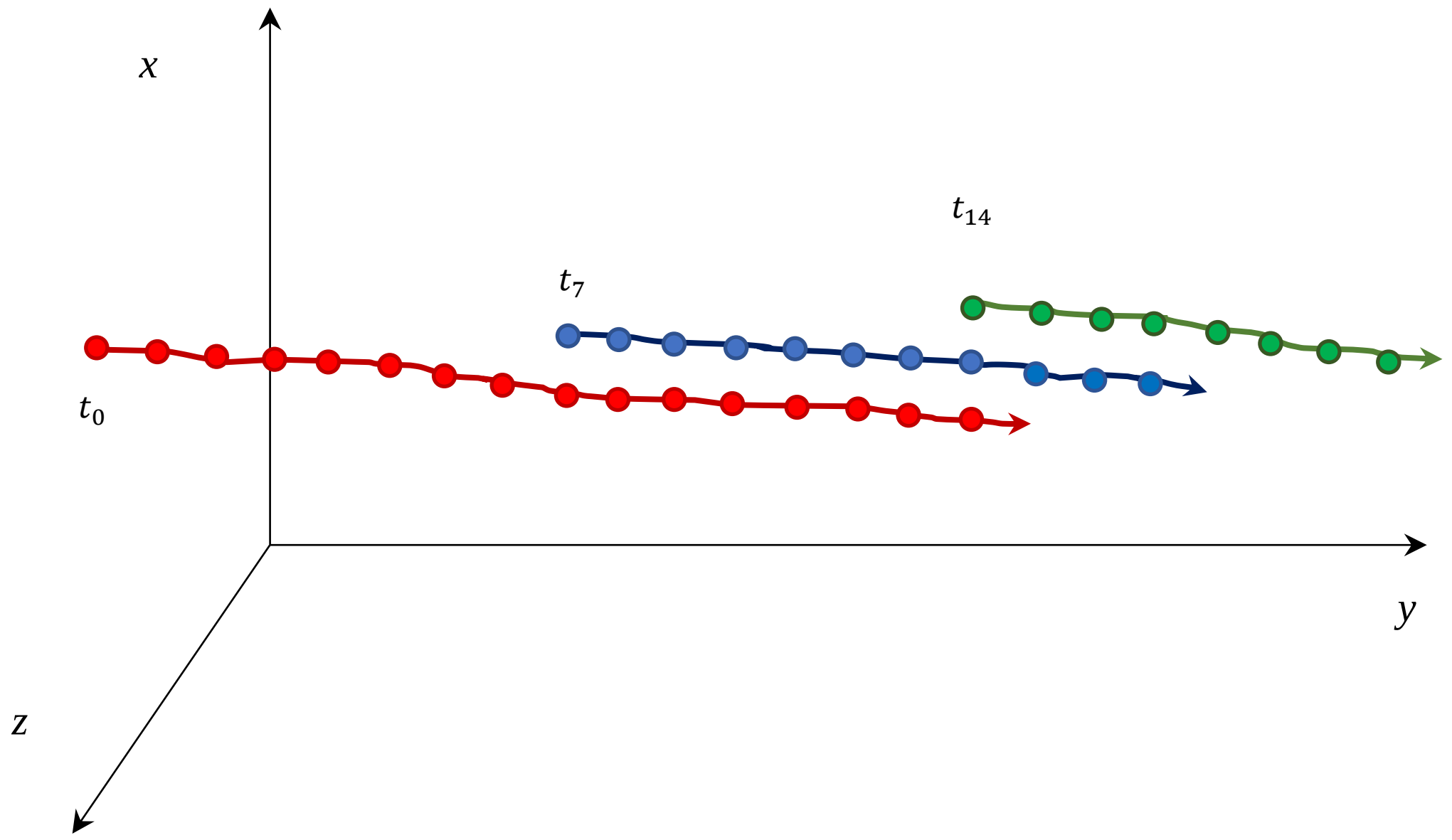

**Figure 10.** Pitches of layers 1-3 illustrated as a pathline of fluid particles.

The inertia of the initial pitch, B, propels the pitch downwards—this becomes apparent in layer 1, after a few instances $t$ in the timeline $T_{layer\ 1} = \{t_0, \cdots, t_{15}\}$. Intuitively, the inertial force thus instigates the recognition of layer 1 moving upwards. At $t_8$, the addition of a new layer, $T_{layer\ 2} = \{t_8, \cdots, t_{23}\}$, causes the initial intersection (or overlapping) of two pathlines, thereby resulting in a density increase[15]. Since density is proportional to pressure, the onset of layer 2 instigates an increase in pressure as well; however, given that layer 2 produces the initial laminar flow, we cannot interpret the sudden increase as any instance of non-uniformity. The first non-uniformity in the flow is caused by layer 3's density increase, but returns to uniformity at $t_{16}$ (fig. 11); this brief event of non-uniformity fluctuates throughout the étude, though the maintenance of the laminar flow is often unaffected. This behaviour of gradually entering layers continues, whereas, further in, the intersections (and thus density) peak at six pathlines.[16] Highlighting the peak of intersection proves to be not only important because it forms the layers, but also to acknowledge the polyphony in an intramusical sense.

---

[14] A classical representation of three-dimensional space.

[15] Recall section 2.2 where we musically converted density to the quantity of harmonic layers.

[16] See mm. 131 for example.

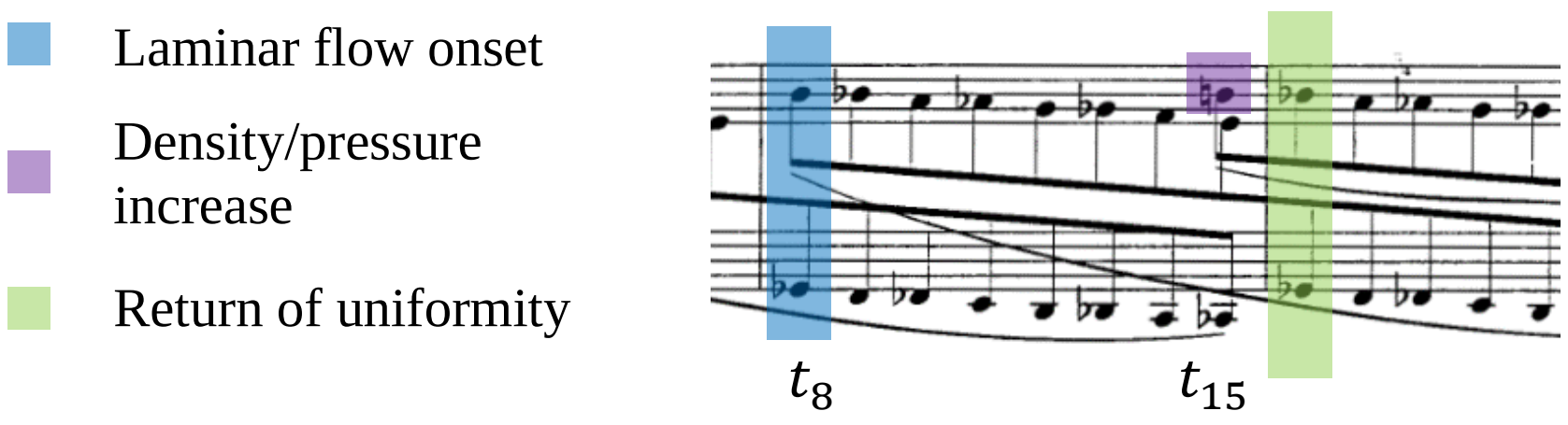

**Figure 11.** Initial increase in pressure and density of the laminar flow.

### 4.2. Mathematically Mapping the Flows, and Fluid Description

A scatter plot[17], often employed in models of fluid dynamics, can successfully capture this motion. Upon collecting pathline data from throughout the score, three scatter plots below demonstrate well the three different fluid states: the initiation of the flow in laminar state (fig. 12), a laminar-turbulent transitional state (fig. 14), and a turbulent state (fig. 15). The X-axis presents the timeline (i.e., each musical interval representative of a single beat), while the Y-axis presents the intervallic activity in the context of registral pitches. For all three scatter plot models, the layers are contained within the scope of six specific octaves such that the Y-axis accounts for octaves above and below the étude's initial octave (C4). To specify the modified Y-axis coordinates, a legend below (table 6) presents the pitch assignments as well as the octaves. Most of the laminar flow layers steadily continue in the range of three octaves, however, turbulent phases yield lower octaves (C3 and C2); for this reason, I have chosen to represent octave 1 and 2 with negative integers to demonstrate the anomaly of the turbulence.

| Y-axis integer | Pitch | Octave |
|---|---|---|
| 47 | B | 6 |
| 46 | B♭ | |
| 45 | A | |
| 44 | A♭ | |
| 43 | G | |
| 42 | G♭ | |
| 41 | F | |
| 40 | E | |
| 39 | E♭ | |

[17] A two-dimensional graph which demonstrates the quantitative relation between two variables: X, characterized by the horizontal measurement, and Y, by the vertical measurement.

<table>
<tr><td>38</td><td>D</td><td rowspan="3"></td></tr>
<tr><td>37</td><td>D♭</td></tr>
<tr><td>36</td><td>C</td></tr>
<tr><td>35</td><td>B</td><td rowspan="12">5</td></tr>
<tr><td>34</td><td>B♭</td></tr>
<tr><td>33</td><td>A</td></tr>
<tr><td>32</td><td>A♭</td></tr>
<tr><td>31</td><td>G</td></tr>
<tr><td>30</td><td>G♭</td></tr>
<tr><td>29</td><td>F</td></tr>
<tr><td>28</td><td>E</td></tr>
<tr><td>27</td><td>E♭</td></tr>
<tr><td>26</td><td>D</td></tr>
<tr><td>25</td><td>D♭</td></tr>
<tr><td>24</td><td>C</td></tr>
<tr><td>23</td><td>B♮</td><td rowspan="12">4</td></tr>
<tr><td>22</td><td>B♭</td></tr>
<tr><td>21</td><td>A♮</td></tr>
<tr><td>20</td><td>A♭</td></tr>
<tr><td>19</td><td>G♮</td></tr>
<tr><td>18</td><td>G♭</td></tr>
<tr><td>17</td><td>F</td></tr>
<tr><td>16</td><td>E♮</td></tr>
<tr><td>15</td><td>E♭</td></tr>
<tr><td>14</td><td>D♮</td></tr>
<tr><td>13</td><td>D♭</td></tr>
<tr><td>12</td><td>C</td></tr>
<tr><td>11</td><td>B</td><td rowspan="12">3</td></tr>
<tr><td>10</td><td>B♭</td></tr>
<tr><td>9</td><td>A</td></tr>
<tr><td>8</td><td>A♭</td></tr>
<tr><td>7</td><td>G</td></tr>
<tr><td>6</td><td>G♭</td></tr>
<tr><td>5</td><td>F</td></tr>
<tr><td>4</td><td>E</td></tr>
<tr><td>3</td><td>E♭</td></tr>
<tr><td>2</td><td>D</td></tr>
<tr><td>1</td><td>D♭</td></tr>
<tr><td>0</td><td>C</td></tr>
</table>

| | | |
|---|---|---|
| -1 | B♮ | 2 |
| -2 | B♭ | |
| -3 | A♮ | |
| -4 | A♭ | |
| -5 | G♮ | |
| -6 | G♭ | |
| -7 | F | |
| -8 | E♮ | |
| -9 | E♭ | |
| -10 | D♮ | |
| -11 | D♭ | |
| -12 | C | |
| -13 | B♮ | 1 |
| -14 | B♭ | |
| -15 | A♮ | |
| -16 | A♭ | |
| -17 | G♮ | |
| -18 | G♭ | |
| -19 | F | |
| -20 | E♮ | |
| -21 | E♭ | |
| -22 | D♮ | |
| -23 | D♭ | |
| -24 | C | |

**Table 6.** Pitch assignments and octave range for Y-axis of flow models.

The first example of laminar activity models the opening measures of the étude as discussed above (4.1). Although this relatively ideal model dissipates at the time of the laminar-turbulent transition phase, the constancy of direction in conjunction to gradual decreases in pressure among respective layers (from $Pa = 13$ to $Pa = 7$) at $t_{497}$ (mm. 63) allows the expectation for a return to a laminar flow—which indeed occurs at $t_{505}$ (mm. 64), where both the direction and pressure remain constant.

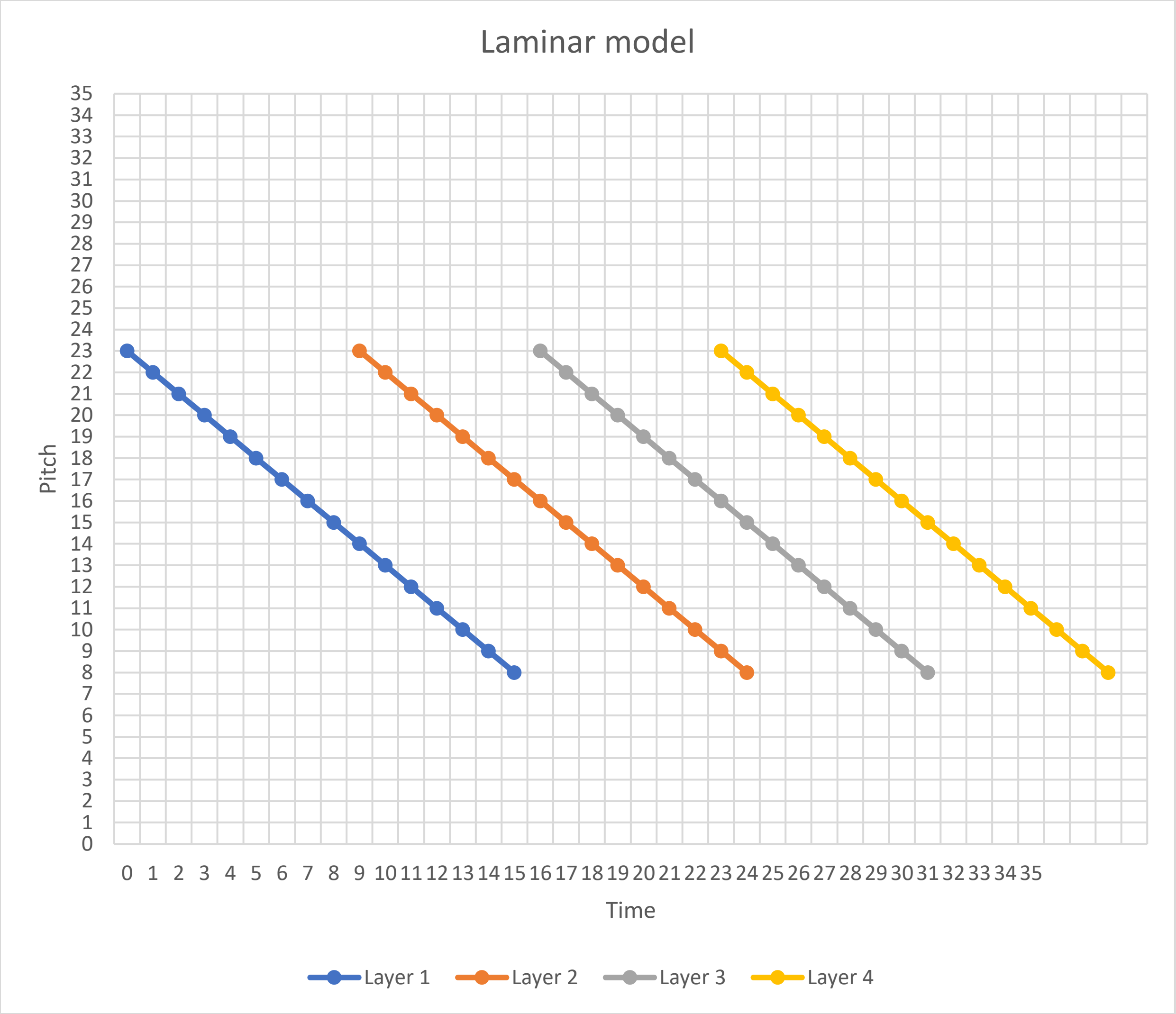

**Figure 12.** Étude No. 9, mm. 1-5. Laminar phase exemplar.

The laminar model formed by the opening measures continues until mm. 25, where a turbulent spot occurs—evidently the initiation of the laminar-turbulent phase in the work. A change in direction and pressure aligns with the transitional phase properties, but not a change in the velocity profile. This poses an important question: if turbulent motion yields a change in both pressure and velocity alongside the occurrence of smooth layer disturbances, can a mere change in pressure and smooth layer disturbances constitute turbulent motion, irrespective of its absence in nature? This constant velocity profile, however, demonstrates the behaviour of a *plug flow*. Accordingly, one could claim that a unique laminar-turbulent flow type is constructed in the musical sense.

Of the four layers in mm. 25 (fig. 13), layer $A$—which we will identify as the lowest layer (i.e., the bass)—includes a pressure change at $t_{198}$ such that $Pa = 6$ increases by 3 to $Pa = 9$. At $t_{198+8}$ (mm. 26), the flow's initial change in direction occurs (see red brackets in mm. 26 of fig. 13). The initial directional change instigates another at $t_{213}$, and both changes correspond to the decay of the kinetic constancy rather than the shift from a downwards to upwards motion. All other layers preserve the laminar properties, despite the slight pressure change marked by "*cresc. poco a poco*," which designates an increase in $Pa$ by at least 1.

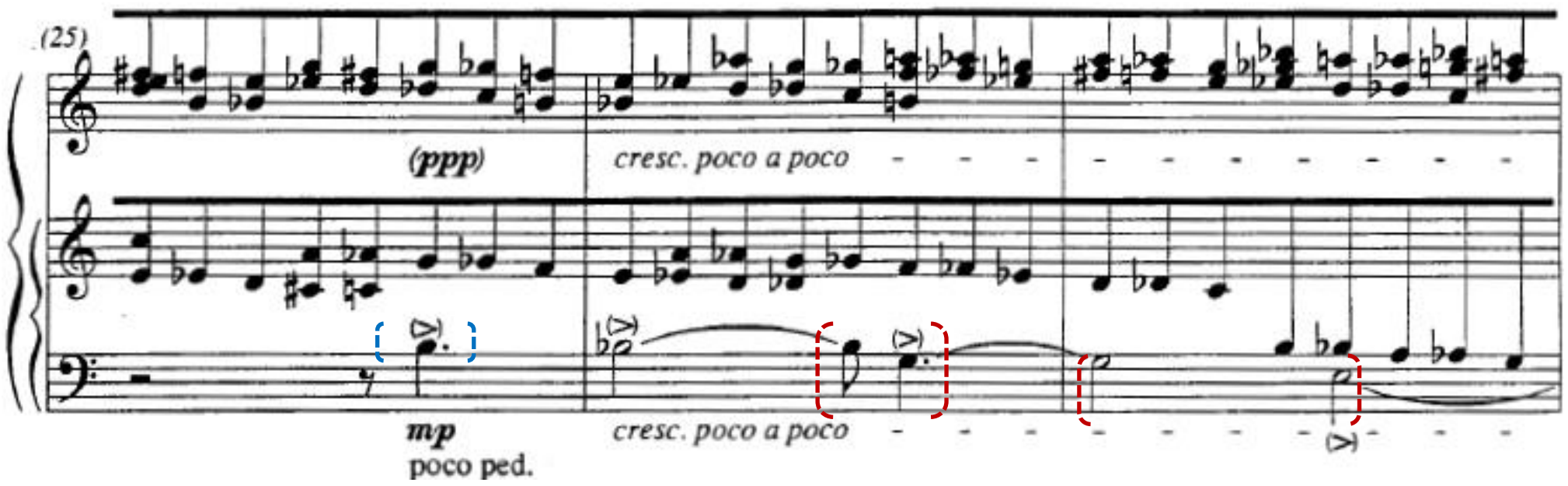

**Figure 13.** Étude No. 9, mm. 25-27. Blue denotes pressure change, and red denotes directional change.

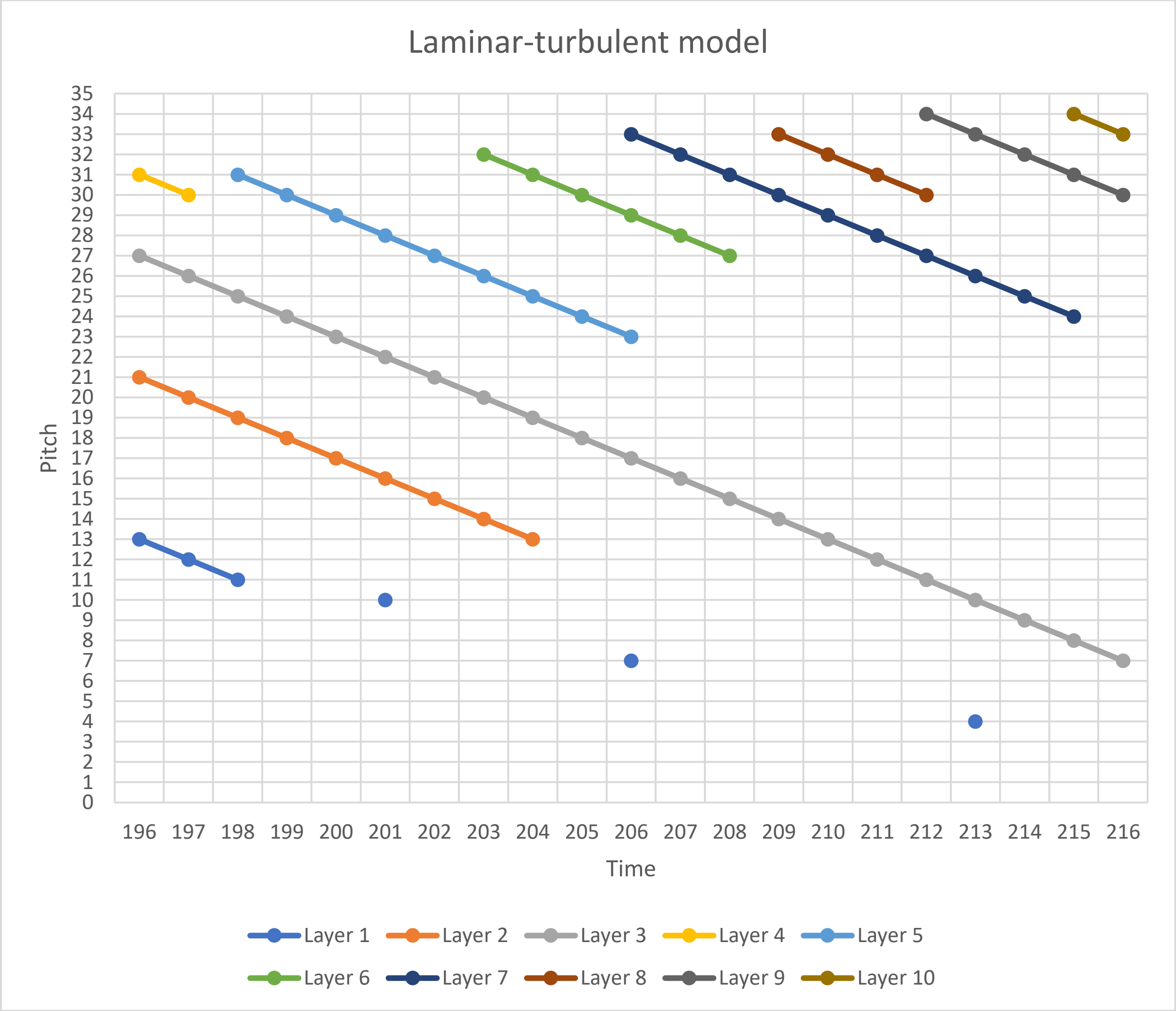

**Figure 14.** Étude No. 9, mm. 25-27. Laminar-turbulent phase exemplar.

Perhaps the best example of a turbulent phase is at $T = \{t_{406}, \cdots, t_{432,}\}$ (fig. 15), where the anomalies in direction and pressure change behave more disorderly between the layers than any other area in the étude. Similar areas at a later time (e.g., mm. 115-125) also yield immense changes in pressure and direction, but our selected timeline presents a paragon in that only two layers are preserved at a time; others contain three or more layers, or less directional changes, and thus resemble more laminar-turbulent transitional phases rather than pure turbulence. Within the chosen $T$, both manifestations of directional change occur; that is, upwards motion, and constancy decays. Naturally, as typical with turbulence, all pressure rates in the musical flow are quite high. Represented as a collection, the pressure range of the timelines holds $T(Pa) = \{10,11,12,13\}$, and, though the range demonstrates a linearity, the pressure varies per layer. Alike the laminar-turbulent phase, no change in velocity profile arises.

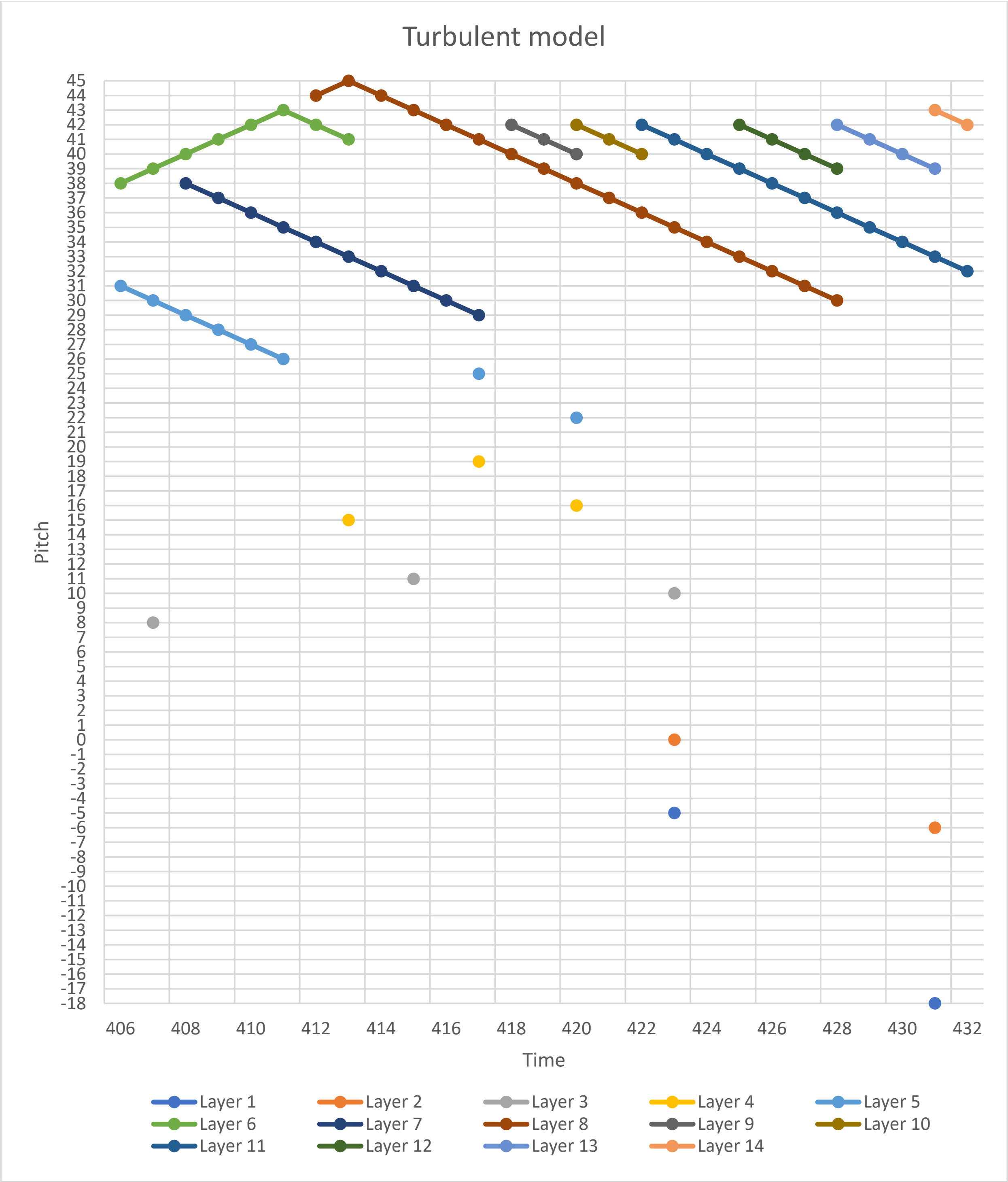

**Figure 15.** Étude No. 9, mm. 51-54. Turbulent phase exemplar.

Surprisingly, the laminar-turbulent transition and turbulent phases do not entail changes in velocity—only in pressure and direction. Despite the unexpected anomaly in the constancy of the velocity profile throughout the étude, one could claim that Ligeti had constructed a unique, controlled[18] flow timeline. Hence, in a musical framework, fidelity is paid to the conventions of natural laminar and turbulent flows at the expense of one characteristic magnitude.

# 5 Conclusion

In this brief study, we proposed a musicological framework to analyze musical structures through the perspective of fluid dynamics with a focus on laminar and turbulent flows. Through designing extramusical methods of conceptual conversions, dynamical representations of pressure rates, adaptive conditions, and scatterplots, a general method of musical analysis was presented. Ligeti's étude no. 9 consists of a long structure that closely corresponds to the proposed musical descriptions of laminar and turbulent flows; however, at the expense of one naturally occurring property. We observed that Ligeti constructed a unique laminar-turbulent timeline, where changes in pressure and direction are focal, but a change in the velocity profile is absent. Although Ligeti's étude no. 9 constitutes a plausible example for the construction of musical laminar and turbulent flows, other similar examples may exist in the literature, thereby securing an incentive for future research into musical flows. This study not only elucidated the plausibility of applying fluid dynamics to music, but also the general approach of explaining musical motion with metaphors. Particularly, a metaphorization granted us a unique capacity to interpret Ligeti's étude (and music in general) differently, as well as an advantageous approach for analytic understanding.

# 6 Appendix of Symbols

$\hat{X}$ scale degree.

$t_n$ timeline *n* (where *n* enumerates a natural number).

$T$ set of timelines.

$p$ pressure magnitude.

[18] In this sense, similar to an engineered construction.